\documentclass[12pt]{article}
\usepackage{graphicx}
\usepackage{epstopdf}
\usepackage{amsmath}
\usepackage{amsfonts}
\usepackage{amssymb}
\usepackage{enumerate}
\usepackage{amsthm}
\usepackage{array}
\usepackage{epsfig}
\usepackage[small,bf]{caption}
\usepackage{appendix}
\usepackage{dsfont}
\usepackage{hyperref}
\def\1ad{\mbox{\normalsize $^1$}}
\def\2ad{\mbox{\normalsize $^2$}}
\def\3ad{\mbox{\normalsize $^3$}}
\def\4ad{\mbox{\normalsize $^4$}}
\def\5ad{\mbox{\normalsize $^5$}}
\def\6ad{\mbox{\normalsize $^6$}}
\def\7ad{\mbox{\normalsize $^7$}}
\def\8ad{\mbox{\normalsize $^8$}}

\def\beq{\begin{equation}}                     %
\def\eeq{\end{equation}}                       %
\def\bea{\begin{eqnarray}}                     
\def\eea{\end{eqnarray}}                       
\def\dj{\hbox{d\kern-0.347em \vrule width 0.3em height 1.252ex depth
-1.21ex \kern 0.051em}}

\def\half{{1\over 2}\,}

\def\ket{\rangle}

\def\pt{\partial}

\def\shalf{{\mbox{$\half$}}}

\def\Dirac{\,\raise.15ex\hbox{/}\mkern-13.5mu D}
\def\dirac{\,\raise.15ex\hbox{/}\kern-.57em \partial}
\def\pslash{\,\raise.15ex\hbox{/}\kern-.57em p}

\begin{document}

                     %

\newcommand{\sheptitle}
{On the spectrum of nonrelativistic AdS/CFT}
\newcommand{\shepauthora}
{{\sc
 Jos\'e L.F.~Barb\'on and Carlos A. Fuertes}}

\newcommand{\shepaddressa}
{\sl
Instituto de F\'{\i}sica Te\'orica  IFT UAM/CSIC \\
 Facultad de Ciencias C-XVI \\
C.U. Cantoblanco, E-28049 Madrid, Spain\\
{\tt jose.barbon@uam.es}, {\tt carlos.fuertes@uam.es} }

\newcommand{\shepabstract}
{ 
\noindent

We develop a Hamiltonian picture for a family of models of nonrelativistic AdS/CFT duality. The Schr\"odinger group is realized via the conformal quantum mechanics of De Alfaro, Fubini and Furlan in the holographic direction. We show that most physical requirements, including the introduction of harmonic traps, can be realized with {\it exact} AdS metrics, but without any need for exotic matter sectors in the bulk dynamics. 
This Hamiltonian picture can be used to compare directly with  many-body spectra of fermions at unitarity on harmonic traps, thereby providing a direct physical interpretation of the holographic radial coordinate for these systems.  Finally, we add some speculations on the dynamical generation of mass gaps in the AdS description,  the resulting quasiparticle spectra, and the analog of `deconfining' phase transitions that may occur.}

\begin{titlepage}
\begin{flushright}
{IFT UAM/CSIC-08-37\\
}

\end{flushright}
\vspace{0.5in}
\vspace{0.5in}
\begin{center}
{\large{\bf \sheptitle}}
\bigskip\bigskip \\ \shepauthora \\ \mbox{} \\ {\it \shepaddressa} \\
\vspace{0.2in}

{\bf Abstract} \bigskip \end{center} \setcounter{page}{0}
 \shepabstract
\vspace{2in}
\begin{flushleft}
\today
\end{flushleft}


\end{titlepage}

\newpage


\setcounter{equation}{0}

\section{Introduction}
\noindent

A proposal for an holographic realization of systems with nonrelativistic conformal symmetry (the so-called Schr\"odinger group)\cite{Hagen,Niederer} was put forward in a couple of interesting recent papers \cite{son, mcg}. 
The natural area of application of these constructions would be that of many-body systems with interacting Hamiltonians of the form 
\beq\label{contt}
{\cal H}_\psi =- \sum_a {1\over 2m_a} 
\psi^\dagger\;{\vec \pt}^{\,2}\, \psi -  \sum_{a,b,c,d} C_{abcd}\;  \psi^\dagger_a \psi^\dagger_b \psi_c \psi_d\;,
\eeq
where the latin indices stand for internal degrees of freedom, such as spin or flavor species and the $C$'s
define a set of running couplings (see also \cite{JackiwPi}). The Schr\"odinger group classifies fixed points, either ultraviolet (UV) or infrared
(IR) of these microscopic Hamiltonians, and characterized by a spectrum of scaling composite operators built from the
elementary fields $\psi_a$ (cf.~\cite{wise,Kaplan}).  

The analysis of nontrivial fixed points may be accessible via  traditional vector-like large $N_f$ limits, in analogy
with the Gross--Neveu or CP$^N$ models. 
For example, one may take flavor indices furnishing an internal $O(N_f)$ or $Sp(2N_f)$ symmetry 
of (\ref{contt}) (cf.~\cite{sachdev,veillette}). In an appropriate  large-$N_f$ limit certain composite operators may become `free',  such as ${\cal O}_{\psi\psi} \sim \sum_{a,b} d_{ab} \psi_a \psi_b$, with $d_{ab}$ an invariant tensor of the flavor group, or operators built from various Noether currents. The operators are free in the sense that three and higher point functions vanish as
positive powers of $1/N_f$ as $N_f \rightarrow \infty$, whereas the anomalous dimensions are finite and calculable in the same limit, through a convenient resummation of Feynman diagrams (see \cite{mehen} for a recent analysis of this sort). These resummations are controlled by
an effective coupling parameter, $c_{\rm eff} = C N_f$ that is kept fixed in the large $N_f$ limit, much like the similar situation with the 't Hooft coupling of large $N$ Yang--Mills theories. In this context, one expects the holographic description to
be a good approximation when $c_{\rm eff} \gg 1$, so that the radius of curvature of the background geometry becomes
small for large $c_{\rm eff}$. 

Such  holographic picture would work similarly to that of relativistic  $O(N_f)$ models, as in Ref.~\cite{klebpol}. According to the standard AdS/CFT rules \cite{classic}, there must be a bulk field $\phi$ acting as a source of
each `free' composite operator ${\cal O}_\phi$. Correlation functions of scaling operators are computed in terms of
boundary Green's functions of $\phi$ with bulk interactions controlled by couplings suppressed by powers of $1/N_f$.  At $N_f =\infty$ the bulk dynamics becomes free, and the mass parameters of $\phi$
control the spectrum of conformal weights $\Delta ({\cal O}_\phi)$. The conformal group itself is realized
as an isometry group of the holographic background. In principle, the kinematical structure can be defined even for finite $N_f$, with no considerations of a large $N_f$ limit \cite{son,mcg},  provided one is prepared to face a strongly coupled problem in the bulk.

In this paper we develop the Hamiltonian picture of the models introduced in Refs.~\cite{son, mcg}. 
In section 2 we begin by reviewing the construction of Refs.~\cite{son, mcg} and point out some  interesting modifications of detail. In section 3 we find the single-particle bulk Hamiltonian of a bulk bosonic field and encounter the De Alfaro, Fubini, Furlan system \cite{daff}. In the process we emphasize that all physical results can be obtained from an exact AdS bulk metric, with no extra matter required. We also  show how to describe the harmonic trapping in this system and point out a fully geometrical interpretation of the procedure that involves analytic continuation of a certain topological AdS black hole.  We end in section 5 with some speculations regarding applications which are naturally suggested by the Hamiltonian methods developed in this
paper. In particular we consider a matching to harmonically trapped fermions at unitarity, as well as
hypothetical patterns of quasiparticle spectra in models with dynamical mass gap generation.   Finally, we end with our conclusions and open questions raised by this work.

While this paper was being prepared for publication, the recent paper   \cite{today} was posted,  containing some overlap with our findings, particularly in
what regards the emphasis on using pure AdS constructions, without any exotic matter sectors supporting  the deformed metrics of \cite{son, mcg}. 

\section{Bulk metrics with Schr\"odinger symmetry}
\noindent

Following \cite{son}, the Schr\"odinger algebra in $d$ spatial dimensions can be conveniently seen as a projection of the full 
relativistic conformal algebra in $d+2$ spacetime dimensions, down to a subgroup that leaves fixed a light-cone momentum
$p^+$. In fact it is not possible to embed the Schr\"odinger group into the conformal group with the same number of
spatial dimensions \cite{burdet,havas}. See also \cite{gomis}.  This suggests that the appropriate metrics can be obtained as a  projection of   an AdS$_{d+3}$ space. Starting from the Poincar\'e patch of AdS$_{d+3}$ spacetime with curvature radius $R$ 
\beq\label{adsm}
ds^2 = {r^2 \over R^2} (-2dx^+ dx^- + d{\vec x}^{\,2} ) + {R^2 \over r^2} dr^2 \;
\eeq 
in an appropriate  light-cone frame, the idea is to project the dynamics onto the sector with fixed eigenvalue of the
light-cone momentum $p^+ = \pt /\pt x^-$. 
 Interpreting now $x^+ =t$ as a time coordinate, we rewrite the previous metric as 
 \beq\label{adsmm}
 ds^2 = {r^2 \over R^2} (-2dt\, d\xi + d{\vec x}^{\,2} ) + {R^2 \over r^2} dr^2 \;
,\eeq
 where we have renamed $x^- = \xi$.  The generator or time translations   $p^- = i\pt_t = H$ is a Hamiltonian, and one realizes the full Schr\"odinger group in terms of those isometries of (\ref{adsmm}) that leave  the $\xi$-momentum $p^+ = -i\pt_\xi = M$  fixed and interpreted as a mass parameter. A quantization of the mass parameter is realized by a compactification of the $x^- = \xi$ direction on a light-like circle.  The resulting isometry group includes the full Galilean group in $d$ spatial dimensions as a subgroup, together with an $SL(2, \mathds{R})$ subgroup generated by the Hamiltonian, $H$, a dilation generator $D$, and a special conformal transformation $C$.   
 
 We will find it convenient to follow the parametrization in \cite{mcg} and consider a more general set of deformations
 of (\ref{adsmm}), labeled by the so-called `dynamical exponent' $z$: \beq\label{sont} ds^2 = -2\gamma^2 {r^{2z} \over
   R^{2z}} dt^2 + {r^2 \over R^2} (-2dt\,d\xi+ d{\vec x}^{\,2} ) + {R^2 \over r^2} dr^2 \;, \eeq where $\gamma^2$ is a
 positive real number. The case $z=2$, with the normalization $\gamma^2=1$ was the main subject of analysis in
 Ref.~\cite{son}.  The isometries of these deformed metrics include the Galilean group in $d$ dimensions, as well as a
 dilation symmetry \beq\label{dilg} (t, {\vec x}, r, \xi) \longrightarrow (\lambda^z \,t, \lambda \,{\vec x},
 \lambda^{-1}\, r, \lambda^{2-z} \,\xi) \;.  \eeq The Schr\"odinger group arises in the particular case $z=2$, and
 includes an additional generator of special conformal transformations.  {\it A priori}, any model with a fixed
 quantized mass spectrum, $M = -i\pt_\xi$, breaks explicitly all $z\neq 2$ dilations, suggesting that only $z=2$
 metrics are physically relevant for the purposes of this paper. However, one of our main results is the recognition
 that the metric (\ref{sont}) with $z=\gamma^2 =1$ is a valid background supporting the full Schr\"odinger group at the
 quantum level, at least in the free approximation of the bulk degrees of freedom.  This special background will have
 peculiar geometrical properties: it is nothing but pure AdS and a suitable generalization of it will be seen to  provide the dual of
 the system placed in a harmonic potential, while remaining pure AdS. That the $z=1$ metric is pure AdS can be seen by starting
 from the AdS metric in Poincar\'e coordinates
    \begin{equation}
      ds^2 = \frac{r^2}{R^2}(-d\tau^2 + d\chi^2) + \frac{R^2}{r^2}dr^2 + \frac{r^2}{R^2} d{\vec x}^{\,2}
    \end{equation}
    and making the change of coordinates
    \begin{equation}
      t = \frac{\chi-\tau}{5^\frac{1}{4}} \quad, \qquad \xi = \frac{-\lambda_+ \chi + \lambda_- \tau}{5^{\frac{1}{4}}}
    \end{equation}
    with $\lambda_\pm = \frac{1}{2}\left(1\pm\sqrt{5}\right)$. Then the metric becomes
 \begin{equation}
      ds^2 = \frac{r^2}{R^2}(-dt^2 - 2 dt d\xi) + \frac{R^2}{r^2}dr^2 + \frac{r^2}{R^2} d{\vec x}^{\,2} \qquad.
    \end{equation}
  
  A second important lesson of the Hamiltonian formalism will be that the value of $\gamma^2$ for $z=2$ metrics is largely irrelevant for physical purposes. Indeed, in practice  one may smoothly set 
   $\gamma=0$  and work with  the exact AdS background (\ref{adsmm}), without the need for exotic compensating matter sectors that would support the deformed metrics with $\gamma^2 > 0$.  Therefore, for the purposes of this paper
   we are mostly interested in the  $z=1=\gamma^2 $ metrics, together with the undeformed pure AdS metrics $\gamma^2 =0$.

\section{Energy spectrum}

\noindent

 We shall model bulk degrees of freedom in terms of a complex, minimally coupled scalar field of mass $m$, in the free approximation, with action  
\beq
\label{kg}
S_\phi = -\shalf \int d^{d+3} x \,\sqrt{-g} \,\left(g^{\mu\nu} \pt_\mu \phi^* \pt_\nu \phi + m^2 \phi^* \phi \right) 
\,.
\eeq

Our reduction procedure will be imposed by compactifying the $\xi$ direction on a circle of radius $1/M$, and subsequently projecting all degrees of freedom onto the sector with fixed $\xi$ momentum $-i\pt_\xi = M$. Alternatively,
we may  compactify on a circle of radius $1/\mu$ and project onto the sector with $\xi$-momentum $M= N \mu$, with a positive  
integer number, $N$,  acquiring the interpretation of a conserved particle number.  In either case, working with (\ref{kg})
the procedure boils down to imposing a restriction to field configurations  of the form\footnote{Dynamical reductions of this type were studied in \cite{duval}.}
\beq\label{fc}
\phi (t, \xi, r, {\vec x}\,) =  e^{iM\xi}  \,\beta (r) \,\varphi(t,  r, {\vec x}\,)
\;,\eeq  
where $\beta(r)$ is  an appropriate rescaling function that we include for convenience. For a metric as general as
\beq\label{genm}
ds^2 = -2A(r, {\vec x}\,) dt^2 -2 B(r) d\xi dt + G(r) dr^2 + F(r)\,d{\vec x}^{\;2}
\;,\eeq
with {\it arbitrary} ${\vec x}$ dependence on the time-time component, insertion of the {\it ansatz} (\ref{fc}) in the action with
\beq\label{defbeta}
\beta(r) = \left(4\pi^2 B(r) F(r)^d \right)^{-1/4}
\eeq
 yields the action of a non-relativistic quantum mechanical system
\beq\label{genac}
S_M= -\int dt \,d^d x \,d\rho\, \left[ \half\left( -i \varphi^* \pt_t \varphi + i \pt_t \varphi^* \varphi\right) + {1\over 2M} \left( |{\vec \pt} \varphi |^2 +
|\pt_\rho \varphi |^2 \right) + U(\rho, {\vec x}\,) \,|\varphi|^2 \right]\;,
\eeq
 where we have defined a new radial coordinate $\rho$ that solves 
\beq\label{tortuga}
d\rho = dr \sqrt{G(r) \over B(r)}\;,
\eeq
and an effective nonrelativistic  effective potential $U$  given by
\beq\label{pote}
U(\rho, {\vec x}\,) = {m^2 \over 2M} B(\rho) + {M } {A(\rho, {\vec x}\,) \over B(\rho)} + {1\over 8M} \pt_\rho^2 \,\log \left(B(\rho) F(\rho)^d \right) + {1\over 32M} \left[\pt_\rho \, \log\left(B(\rho) F(\rho)^d\right) \right]^2 \;.
\eeq

\subsection{Conformal Quantum Mechanics}

\noindent

 We may now turn to the particular case of the family of metrics with Galilean invariance (\ref{sont}) and we find an associated Hamiltonian
\beq\label{hamz}
H={{\vec p}^{\;2} \over 2M} -{1\over 2M} {d^2 \over d\rho^2} + {(d+1)(d+3) + 4(mR)^2 \over 8 M\rho^2} + \gamma^2 {M} \left({R \over \rho}\right)^{2z-2}\;,
\eeq
 where ${\vec p} = -i {\vec \pt}$ is the momentum in the spatial coordinates ${\vec x}$ and $r = R^2 /\rho$. For the Schr\"odinger invariant case, $z=2$, we see that
 the total Hamiltonian consists on a separated free part, for a particle of mass $M$, and an extra term
 coming  from the holographic coordinate and featuring a well-known system, i.e.~the conformal quantum mechanics studied in Ref.~\cite{daff}:
\beq\label{kzd}
H_{z=2} = H_{\vec x} +  H_{\rho} (b_2)\;,
\eeq
with $H_{\vec x} = {\vec p}^{\;2} /2M$ and  $H_{\rho} (b)$  given  by
\beq\label{affb}
H_{\rho} (b) = -{1\over 2M} {d^2 \over d\rho^2} + {b \over 2M \rho^2}\;,
\eeq
parametrized by a dimensionless coupling constant, $b$.\footnote{Notice that the value of the mass in  $H_\rho$ is immaterial, since we can rescale it by a rescaling of $\rho$, a consequence of the conformal character of this system.} For $z=2$ we have
\beq\label{valuedos}
b_2 = {(d+1)(d+3) \over 4} + (m^2 + 2\gamma^2 M^2) R^2 \;.
\eeq
We see explicitly that the effect of $\gamma^2$ reduces to a renormalization of the scalar field mass from $m^2$ to ${\bar m}^2 = m^2 + 2\gamma^2 M^2$. 

At this stage, we meet a surprise, since the case $z=1$, {\it a priori} not enjoying a  manifest Schr\"odinger symmetry, does split in the same manner as the $z=2$ case, up to an additive redefinition of the Hamiltonian. We find
\beq\label{hamuno}
H_{z=1} -\gamma^2 {M} = H_{\vec x} + H_{\rho} (b_1)\;,
\eeq
with
\beq\label{buno}
b_1 = {(d+1)(d+3) \over 4} + (m R)^2\;.
\eeq
Incidentally, it is interesting that the required additive shift of the Hamiltonian is just the rest mass for the particular case $\gamma^2 =1$. 
The crucial property of the conformal Hamiltonian $H_\rho = p_\rho^2 /2M + b/2M\rho^2$ is that
it generates an $SL(2,\mathds{R})$ group when combined with the generator of dilations, $D_\rho =\shalf( \rho \,p_\rho+ p_\rho \rho)$,
and special conformal transformations $C_\rho = \shalf M \rho^2$. 

Since the free Hamiltonian $H_{\vec x}$, together with $D_{\vec x} = \shalf ({\vec x} \cdot {\vec p} + {\vec p} \cdot {\vec x} )$ and $C_{\vec x} = \shalf M |{\vec x}|^2$, generates a commuting $SL(2,\mathds{R})$ group, we find that the full spectrum is acted on by the $SL(2, \mathds{R})$ generated by
$$
H-\gamma^2 M \delta_{z,1} = H_{\vec x} + H_\rho \;, \qquad D = D_{\vec x} + D_\rho \;, \qquad C= C_{\vec x} + C_\rho\;,
$$ 
and both the $z=2$ and the $z=1$ systems exhibit full Schr\"odinger symmetry at the quantum level. The action of the
dilation of the quantum
$SL(2,\mathds{R})$ group on the coordinates corresponds to (\ref{dilg}) with $z=2$. However this dilation transformation {\it is not} an isometry of the $z=1$ metric, so that we cannot expect it to survive
local interactions in the bulk. We will return to this issue at the end of section \ref{sec:3.3}.

  In both situations the dynamics of the holographic direction contribute a continuous spectrum of excitations which are well-contained in the UV regime ($r\rightarrow \infty$ or $\rho \rightarrow 0$) since the potential diverges there, and accumulate at the IR end, as the conformal potential vanishes  in the 
$r\rightarrow 0$ limit.  This situation is analogous to the  behavior of
a relativistic conformal field theory in infinite volume, as obtained from the study of bulk dynamics on the Poincar\'e patch of AdS. 

The form of the radial potential suggests that the appropriate UV/IR correspondence in this model is
$\omega \sim 1/M\rho^2$, where $\rho$ is the radial variable appearing in (\ref{affb}) and $\omega$ is a physical quantum of energy. 
If we formally impose an infrared cutoff in the system, say at $\rho \sim L_\rho$, the spectrum at high energies will discretize in momentum modes $p_\rho \sim 2\pi n_\rho /L_\rho$. This means that the high-energy density of states
gets a contribution from the holographic direction as an extra dimension of length $L_\rho$. We will dwell on the significance of this fact at the end of this paper.

For $z<1$, the non-conformal term in the effective potential of (\ref{hamz})  grows at large $\rho$ and sets a mass gap in the system, in analogy with similar deformations in the so-called AdS/QCD models.

\subsection{Trapping the system}

\noindent

A standard strategy at discretizing the spectrum in standard AdS/CFT models is to put the system on a finite volume. If the procedure is carefully chosen, the resulting Hamiltonian may still enjoy useful constraints from the conformal symmetry. In standard AdS/CFT models, this happens when putting the CFT on a spatial sphere of constant curvature, whose Hamiltonian is proportional to the dilation operator in flat Minkowski space, hence one can equate energies on the sphere to conformal dimensions of local operators on the hyperplane \cite{mack}. 

There is an analogous construction in systems with non-relativistic conformal symmetry \cite{sonnishida}. Confining the system 
to  a harmonic trap of frequency $\Omega$ corresponds to adding the potential 
$$
V_{\rm trap} =  \half M \Omega^2 |{\vec x} |^2\;.
$$
Hence, from the explicit form of the generator of special conformal transformations for a Schr\"odinger invariant system we learn that
\beq\label{traph}
H_{\rm trapped} = H_{\rm untrapped} + \Omega^2 \,C\;.
\eeq
 The $SL(2, \mathds{R})$ group can then be used to obtain exact 
 properties  of the spectrum such as its discrete gap and a number of 
virial theorems for certain expectation values (cf.~Ref.~\cite{castin}). Furthermore, it can be shown that the spectrum
of $H_{\rm trapped}$  sets the spectrum of conformal weights of local operators in the untrapped system, i.e.~for each
local conformal operator in the untrapped system we have an energy eigenstate of the harmonically trapped system, $|{\cal O}\ket$, whose energy satisfies
\beq\label{theo}
E_{\cal O} =  \Omega \Delta_{\cal O}\;,
\eeq
with $\Delta_{\cal O}$ the conformal weight (eigenvalue of the dilation operator) of the scaling operator ${\cal O}$ (see \cite{sonnishida}). For the model induced by the non-relativistic AdS/CFT
correspondence studied here,  the full generator of special conformal transformations can be written as
\beq\label{fullc}
C = C_{\vec x} + C_\rho = \half M |{\vec x}\,|^{\;2} + \half M \rho^2\;,
\eeq
so that
\beq\label{tototr}
H_{\rm trapped} = {{\vec p}^{\;2} \over 2M} + \half M \Omega^2 |{\vec x}|^2 + H_\rho (b) + \half M \Omega^2 \rho^2\;,
\eeq
 and we see that the prescription (\ref{traph}) induces naturally the corresponding `trapping' in the holographic coordinate $\rho$, with the same characteristic frequency. Notice however that a rescaling of the radial coordinate $\rho \rightarrow \lambda \rho$  induces an effectively rescaled mass $M^2_\lambda = \lambda^2 M$ and consequently the trapping in the holographic variable corresponds to a potential $\shalf M_\lambda^2 \Omega^2 \rho^2$. Hence, we can still change the effective mass of
the radial Hamiltonian without changing the trapping frequency.

It remains to show how to induce the harmonic trapping at the level of the geometrical description. Going back to (\ref{pote}) and setting $B(r) =  F(r) = 1/G(r) = r^2 / R^2 = R^2 /\rho^2$, we may now use the fact
that all the ${\vec x}$ dependence in the effective potential $U({\vec x}, \rho)$ is controlled by the
value of the time-time component of the metric $A({\vec x}, \rho)$. In the absence of an explicit  `external' potential
we have $A(r) = \gamma^2 (r/R)^{2z}$ and, adding the perturbation
\beq\label{potpert}
\delta A({\vec x}, r) = {r^2 \over R^2} \;{1\over M} \delta U({\vec x}, \rho) 
\;,
\eeq
generates a potential  term on the effective Hamiltonian  $\delta U ({\vec x}, \rho)$. The dynamics of the `radial problem'
remains decoupled from the standard spacetime degrees of freedom for all potential deformations of the form
\beq\label{defs}
\delta U({\vec x}, \rho) = V({\vec x}\,) + v(\rho)\;,
\eeq
with otherwise arbitrary functions $V$ and $v$. The  
$SL(2, \mathds{R})$ covariant harmonic trap is induced by quadratic deformations with the same coefficient: $V({\vec x}\,) = \shalf M\Omega^2 |{\vec x}\,|^2$ and $v(\rho) = \shalf M\Omega^2 \rho^2$.

Harmonic trapping is a rather realistic situation in actual experiments involving cold atoms \cite{castinrev,giorgini}. From the theoretical point of view it has the additional interest of being exactly solvable. Indeed, the Hamiltonian
\beq\label{conftr}
H_{\rm conformal\;trapped} = -{1\over 2M} {d^2 \over d\rho^2} + {b \over 2M \rho^2} + \half M \Omega^2  \rho^2
\eeq
is diagonalized by eigenfunctions
\beq\label{lag}
f_n (\rho) = \rho^{\nu +\half} L_n^\nu (\sqrt{2} M \Omega \rho^2) \,e^{-M\Omega \rho^2 /\sqrt{2}}\;,
\eeq
where $\nu$ is the real positive solution of 
\beq\label{defnu}
\nu^2 = b + {1\over 4}
\eeq
and $L_n^\nu$ is a generalized Laguerre polynomial. The corresponding spectrum of eigenvalues is
\beq\label{specrho}
\varepsilon_q = \Omega (\nu + 1 + 2q)\;, \qquad q\in \mathds{Z}_+\;.
\eeq
These formulae assume that $\nu^2 >1$. For $0<  \nu^2 \leq 1$, corresponding to $-1/4 < b \leq 3/4$ and including some cases of attractive potentials for negative $b$, there are two possible normalizable solutions at $\rho =0$. One is the
function in (\ref{lag}), and the other is the same function with $\nu \rightarrow -\nu$.  The spectrum constructed on top of this second branch of solutions is given by (\ref{specrho}) with the same redefinition $\nu \rightarrow -\nu$. \footnote{The solutions   with asymptotics $\rho^{\half-\nu}$ are  sometimes referred to as `resonant'.   Even if they are normalizable on $\rho \in [0, +\infty)$ for $0< \nu^2 <1$, their first derivative is not. Hence, the kinetic energy has divergent expectation value in this branch of solutions, as has the potential energy, the two divergences cancelling out to give a finite total energy.}

These two possible quantizations of
the system are the nonrelativistic incarnation of a well known instance in the standard AdS/CFT correspondence \cite{wittenl, magoo}. Going
back to the bulk parameters, the two inequivalent quantizations correspond to  the range in effective masses
\beq\label{rangm}
-\left({d+2 \over 2}\right)^2 < ({\overline m} R)^2 \leq 1- \left({d+2 \over 2} \right)^2\;,
\eeq
where ${\overline m}^2 = m^2 + 2\gamma^2 M^2 \delta_{z,2}$.  The lower limit on ${\overline m}^2$ allows for tachyonic bulk fields
in a finite range of masses. This tachyonic bound coincides exactly with the Breitenlohner--Freedman bound of ${\rm AdS}_{d+3}$ (cf.~\cite{bfbound}). Since we define the nonrelativistic AdS/CFT correspondence as a reduction of the higher-dimensional AdS/CFT map, we find this result rather satisfying.  Violating the Breitenlohner--Freedman bound in this context, i.e.~continuing to $b< -1/4$, makes the conformal potential too strongly attractive and the Hamiltonian problem is not selfadjoint (cf.~\cite{galindo}). 

The full spectrum of $H_{\rm trapped}$ is given by
\beq\label{flls}
E_{{\vec n}, q}^\pm = \Omega \left({d \over 2} +  \sum_{i=1}^d n_i + 2q + 1 \pm \nu\right)\;,\qquad n_i , q \in \mathds{Z}_+
\eeq
where the term $d/2$ comes from the zero-point energy of the $d$-dimensional oscillator in the ${\vec x}$ coordinates and the $(-)$ sign only applies for $0<\nu^2 < 1$.
Using now the general rule (\ref{theo}) we have a spectrum of operator dimensions
\beq\label{opdim}
\Delta_{{\vec n}, q}^\pm =    \sum_{i=1}^d n_i + 2q + {d+2 \over 2} \pm  \nu\;,\qquad n_i , q \in \mathds{Z}_+\;,
\eeq
with the ${\vec n} = q =0$ cases corresponding to the conformal primaries studied in \cite{son, mcg} in terms of the
two-point functions. In particular, the two branches of quantum states in the Hamiltonian formalism, given by
the analytic continuation $\nu \rightarrow -\nu$ for $0< \nu^2 < 1$ correspond to the two branches of boundary conditions 
studied in \cite{son}, with the associated conformal weights $
\Delta^\pm = \Delta(
\pm \nu)$. 

\subsection{A purely geometrical interpretation of the harmonic trapping} 
\label{sec:3.3}
\noindent

In this section we provide a very interesting geometrical interpretation of  the special case $z=1$, which shows full Schr\"odinger  symmetry at the quantum level. It turns out that the $z=1$ metric, deformed with harmonic trapping,  $V({\vec x}\,) =\shalf  M\Omega^2 |{\vec x}\,|^2$, $v(\rho) = \shalf M\Omega^2 \rho^2$, can be written as
\beq\label{rotbh}
ds^2 = -\left({r'^{\,2} \over R^2} + {a^2 (r'^{\,2} - a^2) \over R^6} |{\vec x}\,|^2 \right) dt'^{\,2} - 2 {r'^{\,2} - a^2 \over R^2} d\xi' dt' + {R^2 r'^{\,2} dr'^{\,2} \over (r'^{\,2} - a^2)^2} + {r'^{\,2} - a^2 \over R^2} d{\vec x}^{\;2} \;,
\eeq
under the change of variables $\Omega^2 = 2a^2/ R^4$,  $r^2 = r'^2 -a^2$ and $t=\sqrt{2} t'$, $\xi' = \sqrt{2} \xi$.  After a further double Wick rotation, $\xi' \rightarrow it_B$ and $t' \rightarrow i \theta_B$, the resulting metric  is {\it locally} equivalent to
that of a rotating, extremal, topological AdS black hole, studied in Ref.~\cite{banados}. Of course, the same applies to the non-rotating solution $a=0$, corresponding to the untrapped $z=1$ background.  Since these black holes are topological,
the metric is locally pure AdS spacetime,  and thus solves Einstein's equations with a simple cosmological constant term, without any need for exotic matter that would support the NRAdS/CFT metric, as was required in  Refs.~\cite{son, mcg}. 

It is also remarkable the fact that the correct  $SL(2, \mathds{R})$ covariant harmonic trapping with one and the same frequency on ${\vec x}$ variables and holographic variables arises as a result of the `rotating black hole' prescription. We regard this as  analogous to a well known fact in the standard AdS/CFT correspondence, where the bulk metric associated to the CFT on a round sphere is given by the same AdS spacetime in global coordinates, rather than restricted to the Poincar\'e patch.

Together with the realization that the physical impact of $\gamma^2$ seems rather trivial in the $z=2$ models, this  result suggests that a purely AdS construction is perhaps a more economical and versatile approach to the NRAdS/CFT duality, making it, quite literally, into an
AdS/NRCFT duality. Notice however that the double Wick rotation makes the black hole interpretation rather formal. In addition,  the global identifications implicit in (\ref{rotbh}) are
completely different to those that define the topological black hole.

One of the advantages of the holographic descriptions lies in their specification of  an {\it ansatz} for the interactions that respect the relevant conformal symmetry. In particular, in the context of a bulk scalar degree of freedom we can consider interaction terms
of the form
\beq\label{bulki}
S_{\rm int} \sim  g_k \int d^{d+3} x\;\sqrt{-g}\;   \; |\,\phi\,|^{2k}
\;,
\eeq
or straightforward local generalizations with covariant derivatives. These interaction terms can be used to
compute connected $n$-point functions of those operators ${\cal O}_\phi$ dual to the scalar field $\phi$. Any such
interaction term, written as a perturbation around the ${\rm AdS}_{d+3}$ background metric, respects the conformal
symmetry provided this symmetry is realized as an isometry of the bulk metric.  In this respect, we find a distinction between
the $z=2$ and $z=1$ cases discussed so far. The quantum $SL(2,\mathds{R})$ algebra of the $z=1$ system  is not realized as an isometry of the $z=1$ deformed metric \footnote{This metric, being locally AdS, does have a large isometry group, at least locally,  the point being  that the {\it particular} $SL(2,\mathds{R})$ group that is generated quantum mechanically is not part of it.} so that  generic bulk interactions of the type (\ref{bulki}) will break the
Schr\"odinger group of the pure AdS construction. 

The appropriate interpretation of the pure AdS constructions with $z=1$  is then that of {\it approximate} nonrelativistic fixed
points, i.e.~to the extent that the interactions (\ref{bulki}) are `small' the Schr\"odinger symmetry will classify two
point functions (the spectrum of scaling dimensions) but will be weakly broken at the level of three-point functions. A
typical system with such a behavior is a microscopic model with a vectorlike large $N_f$ limit. In such class of models, the effective couplings satisfy a scaling law $g_k \sim 1/N_f^k$ and one can envisage situations with a fixed point at $N_f =\infty$, broken by $1/N_f$ corrections. Conversely, any fixed point that is exact to all orders of the $1/N_f$ expansion will require use of a background metric of type $z=2$.

\section{Towards applied AdS/NRCFT}

\noindent

In this final section we speculate on  two possible applications of the previous formalism. We comment on broad conceptual lines rather than the precise implementation of the program, that is left for future work.

\subsection{Matching to fermions at unitarity}

\noindent

Fermions at unitarity \cite{fermions} are widely regarded as one of the main arenas of potential applications of these constructions. In
this section we show that a rather precise correspondence may be achieved between the bound state problem of unitarity
fermions on a harmonic trap, and the effective quantum mechanical system described in the preceding sections.

Following Ref.~\cite{castin}, the bound state problem of $N$ particles of mass $\mu$  on a harmonic trap of frequency $\Omega$ and satisfying the unitarity boundary conditions,
\beq\label{cont}
\lim_{{\vec x}_i \to {\vec x}_j} \psi({\vec x}_1, \cdots, {\vec x}_N) = {c_{ij} \over |{\vec x}_i - {\vec x}_j|^{d-2}} + O({\vec x}_i - {\vec x}_j)
\eeq
  can be separated into a center of mass degree of freedom with coordinate
${\vec x} = N^{-1} \sum_i {\vec x}_i$, mass $M = N\mu$,  and free dynamics
\beq\label{freecm}
H_{\rm CM} = -{{\vec \pt}^{\,2} \over 2M} + \half M \Omega^2 |{\vec x}\,|^2\;,
\eeq
plus an `internal' Hamiltonian  problem 
\beq\label{inth}
H_{\rm int} = {1\over 2\mu} \left(-{d^2 \over d\rho^2} + {dN-d-1 \over \rho} {d \over d\rho}+ {\Lambda_{(N,d)} \over \rho^2 }\right) + \half \mu \,\Omega^2 \rho^2\;,
\eeq
in terms of a collective radial coordinate measuring the effective size of the trapped droplet: 
\beq\label{collr}
\rho = \left(\sum_{i=1}^N ({\vec x}_i - {\vec x}\,)^2\right)^{1/2}
\;.
\eeq
In (\ref{inth}), the quantities $\Lambda_{(N,d)}$ are  the eigenvalues of the Laplacian on a proper submanifold of the sphere, ${\bf S}^{dN-d-1}$, related to the relative angular variables, and with complicated boundary conditions. The eigenvalues $\Lambda_{(N,d)}$
encode all the information on the contact constraints (\ref{cont}) and therefore characterize those dynamical details of the system which  do not follow simply from the symmetries. 

By a simple wave function rescaling, this system can be rewritten as the sum of a harmonically trapped  free Hamiltonian of mass $M$ plus
a harmonically trapped conformal quantum mechanics of mass $\mu$ and  coupling parameter
\beq\label{param}
b = \nu^2 - {1\over 4} \;,
\eeq
with 
 \beq\label{matb}
 \nu^2 = \Lambda_{(N,d)} + \left({dN -d-2 \over 2}\right)^2\;.
 \eeq
 The resulting spectrum is known and takes the form 
 \beq\label{spece}
 \varepsilon_q = \Omega (\nu + 1 + 2q)\;, \qquad q \in \mathds{Z}_+\;
 \eeq
 on the internal degrees of freedom. 
 We can match this spectrum by a nonrelativistic AdS construction with radius of curvature $R=1/M$ and  a scalar field of bulk mass $m$ determined by 
 \beq\label{nup}
\nu^2 = \left({d+2 \over 2}\right)^2 + ({\bar m}R)^2  = \Lambda_{(N,d)} + \left({dN -d-2 \over 2}\right)^2
\;,\eeq
 where ${\bar m}^2 = m^2$ for the $z=1$ model,  or ${\bar m}^2 = m^2 + 2\gamma^2 M^2$ for the standard $z=2$ background metrics. Of course, in the case
 $z=1$ we must discard  the additive constant $\gamma^2 M$ to the Hamiltonian.  
  
We see that the crucially unknown parameter of the problem, the spectrum $\Lambda_{(N,d)}$, gets mapped under the
AdS/NRCFT correspondence to the local data of the bulk degrees of freedom, in this case the mass spectrum of bulk fields, according to the precise correspondence (\ref{nup}). In this respect, it is of great interest that the AdS prescription
imposes a lower bound on $\Lambda_{(N, d)}$ following from the Breitenlohner--Freedman bound:
\beq\label{bnub}
\Lambda_{(N,d)} > -\left({dN-d-2 \over 2}\right)^2\;.
\eeq

Finally,  in this construction we obtain a rather direct physical interpretation of the holographic radial coordinate
$\rho$,  as a mean-square radius of the droplet of cold atoms, according to (\ref{collr}). The holographic direction
accounts for the collective excitations of the droplet. 

\subsection{Quasiparticles}

\noindent

We have described the spectral properties of AdS/NRCFT systems either in free space or trapped by appropriate harmonic potentials. In this section we comment on the more general structures that can be expected when the
conformal symmetry is broken dynamically by infrared effects. More specifically, we consider metric deformations
which respect the Schr\"odinger group asymptotically as $\rho \rightarrow 0$ (the UV regime) but induce a mass gap
at some low scale. The situation is analogous to that of harmonic trapping, but restricted to the radial holographic dynamics, i.e.~we keep the system untrapped in the ${\vec x}$ coordinates. 

A natural candidate for such a dynamical mass generation is that of $d=2$ systems with attractive contact interactions, i.e.~systems with delta-function interactions. In this case the quantum-field theory description of bound states involves
a classically scale invariant four-fermion interaction $\psi^\dagger \psi^\dagger \psi \psi$, which undergoes logarithmic
running at one loop. Energies of bound states are formally similar to $\Lambda_{\rm QCD}$ in the sense that they
are non-perturbative in the bare coupling of the delta-function potential. Thus, this is a natural arena to investigate the
geometrical realization of the mass gap. 

\begin{figure}
  \begin{center}
    \epsfig{file=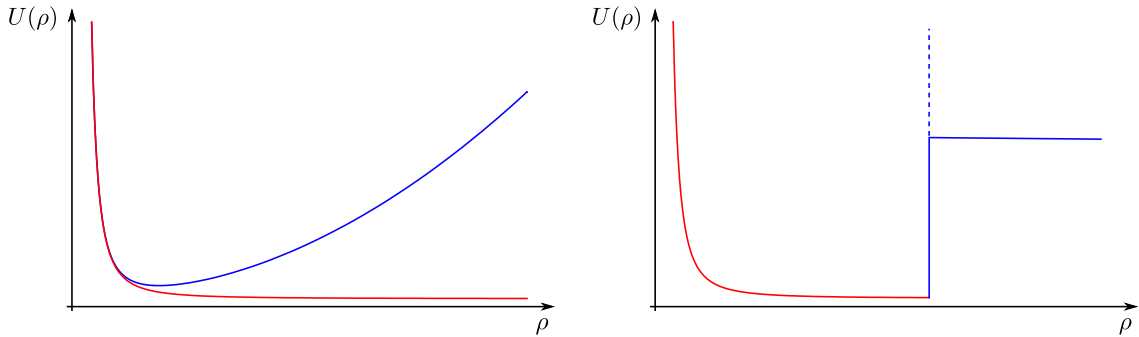, width=15cm}
    \parbox{14truecm}{\caption{Plots of the potential $U(\rho) = \frac{b}{2M\rho^2} + v(\rho)$ that one can induce on
        the holographic radial coordinate $\rho$ to model the generation of a mass gap. The red line presents the conformal potential,  with
        vanishing $v(\rho)$,  while the blue line represents the potential with non-zero $v(\rho)$. On the left plot we have  the harmonic deformation, $v(\rho) =
        \frac{1}{2} M \Omega_\rho^2 \rho^2$,  while on the right we add a sharp wall of finite or infinite height. }
    \label{fig:potential}}
  \end{center}
\end{figure}

While waiting for a better formulated model, there  are two simple phenomenological models for such mass-gap generation: a hard wall at some $\rho= L_\rho$,
imposing a vanishing Dirichlet condition on wave functions at the hard wall, or a harmonic wall 
\beq\label{hwal}
v(\rho) = \half M \Omega_\rho^2 \,\rho^2
\eeq
for which all of our previous results apply, except that the ${\vec x}$ system is not trapped and $SL(2, \mathds{R})$ covariance is broken at the level of the full spectrum (for instance, one loses the theorem (\ref{theo}), but maintains
the exact results on the $\rho$-sector).

Expanding the dynamical field
$$
\varphi(t, {\vec x}, \rho) = \sum_\alpha f_\alpha (\rho) \Psi_\alpha (t, {\vec x}\,)
$$
 in a complete basis of normalized eigenfunctions of the radial Hamiltonian
\beq\label{radh}
H_\rho f_\alpha (\rho) = \varepsilon_\alpha \,f_\alpha (\rho)\;,
\eeq
we can now quantize the system regarding the bulk field as a local operator, leading to the second-quantized non-relativistic Hamiltonian \beq\label{sec}
{\cal H}_{\rm free} = \int d^d x \sum_\alpha \Psi^\dagger_\alpha ({\vec x}\,) \left(-{{\vec \pt}^{\;2} \over 2M}  + \varepsilon_\alpha \right)\, \Psi_\alpha ({\vec x}\,) \;.
\eeq
When the eigenvalue problem (\ref{radh}) has  a discrete spectrum, the index $\alpha$ can be interpreted as a species index for a tower of `quasiparticle types' in (\ref{sec}), with $\varepsilon_\alpha$ acquiring the interpretation of   `internal energies'  for each particle type. This means that these quasiparticle types are to be interpreted as bound states,  {\it molecules}, in  complete analogy with well-known  relativistic constructions in which, for instance, the eigenvalues of the bulk radial problem transmute into a tower of glueball masses, as viewed from the boundary (cf.~\cite{wittenglue}). 

The spectrum of internal energies is determined by the gap potential. In the case (\ref{hwal}) the spectrum of internal
energies is evenly spaced, with step $2\Omega_\rho$, as in (\ref{specrho}) and (\ref{spece}). For a hard wall, the spectrum
of internal energies has a high-energy asymptotics of the form $\varepsilon_\alpha \sim 4\pi^2 \alpha^2 / 2ML_\rho^2$,
with $\alpha$ a positive integer. 

Other situations can be contemplated. For instance, if the gap potential $v(\rho)$ has a minimun at $\rho=0$ but
asymptotes to a positive constant $v_\infty$ as $\rho \rightarrow \infty$, we have a discrete spectrum of a finite number
of quasiparticles with internal energies $0< \varepsilon_\alpha < v_\infty$ followed at higher energies by a continuum
spectrum, as in the Schr\"odinger invariant model (see Fig. 1). 

One interesting piece of information that can be obtained from the knowledge of the internal energies $\varepsilon_\alpha$ is the high-energy density of states. Assume for example that $\Psi^\dagger_\alpha$ in (\ref{sec}) create bosons of internal energy $\varepsilon_\alpha$, then the free energy of such a gas is given by
\beq\label{freee}
\beta F(\beta) = V \sum_\alpha \int {d^d p \over (2\pi)^d} \log\left[1-e^{-\beta\left({{\vec p}^{\,2} \over 2M} + \varepsilon_\alpha\right)}\right]\;.
\eeq
If we simulate the mass gap by a sharp wall at $\rho = L_\rho$, the internal energy spectrum is given asymptotically by $\varepsilon_\alpha =
(2\pi\alpha)^2 / 2ML_\rho^2$, with $\alpha$ a large integer number. Then, the entropy will scale at high temperatures as
\beq\label{scalen}
S(\beta) \sim V L_\rho (MT)^{d+1 \over 2}\;,
\eeq
that is, the tower of quasiparticles contributes like an effective extra  spatial dimension of size $L_\rho$. On the other hand, if we use a harmonic trap of frequency $\Omega_\rho$, the contribution of the quasiparticles sum in (\ref{freee}) amounts to a factor of $T/\Omega_\rho$, or equivalently to {\it two} extra dimensions of effective length $L_{\rm eff} = (M\Omega_\rho)^{-1/2}$. 

Local interaction terms of type (\ref{bulki}) induce corresponding local interactions between the quasiparticles.  In particular, inserting the basic ansatz (\ref{fc}) into (\ref{bulki}) and performing the $\xi$ and $\rho$ integrations we find
effective interaction terms in the effective many body Hamiltonian of the `molecules':
\beq\label{intham}
{\cal H}_{\rm int}^{(k)} \sim {g_k \over M} \sum_{\alpha, \cdots, \beta,
\cdots} V_{\alpha_1, \cdots, \alpha_k,  \beta_1, \dots\beta_k}\int d^d x\; \Psi_{\alpha_1}^\dagger ({\vec x}\,) \cdots  \Psi_{\alpha_k}^\dagger ({\vec x}\,) \Psi_{\beta_1} ({\vec x}\,) \cdots \Psi_{\beta_k} ({\vec x}\,)\;,
\eeq
where the contact interactions are determined by overlapping integrals 
\beq\label{vertice}
V_{\alpha_1, \cdots\alpha_k , \beta_1, \cdots\beta_k } = {1\over 2\pi} \int d\rho  \left({\rho \over R} \right)^{d-1} f_{\alpha_1}^* (\rho) \cdots  f_{\alpha_k}^* (\rho) f_{\beta_1} (\rho) \cdots f_{\beta_k} (\rho)\;.
\eeq

It would be interesting to delve further into the possible applications of this formalism.

\section{Concluding remarks}

\noindent

In this work we have implemented the elements of the Hamiltonian picture for the proposed AdS/RNCFT correspondence in Refs.~\cite{son, mcg}. At the level of free bulk dynamics, we find that the conformal symmetry is realized by a decoupled system in the radial holographic coordinate, given by the conformal quantum mechanics of De Alfaro, Fubini and Furlan. 

This analysis throws two important lessons. The first is the existence of  background metrics that realize the Schr\"odinger group
at the quantum level, even in the absence of the corresponding isometries. The second is that the deformations away
from pure AdS metrics have rather mild effects in the free bulk approximation, such as a simple mass renormalization of scalar fields. This is certainly a welcome state of affairs, because the matter systems suggested in \cite{son, mcg}) as supporters of the relevant metrics look rather exotic. It would be interesting to study in a systematic way whether the mild effects of the $\gamma^2$
parameter in (\ref{sont}) stay so `mild' at the level of interactions, or when the bulk system includes other types of matter, 
such higher spin modes. For example, contact interactions of type (\ref{bulki}) are
independent of $\gamma^2$, since this parameter drops from the expression of the volume density $\sqrt{-g}$. However,
more complicated interactions with derivatives will certainly be sensitive to $\gamma^2$ and the question is whether the
$\gamma^2 \rightarrow 0$ limit is smooth, so that we can use pure AdS backgrounds to describe exact fixed points. Working with a light-like compactification certainly calls for caution, regarding possibly singular quantum effects associated to zero modes. The fact that one explicitly projects the theory
onto a sector of {\it non vanishing} momentum is perhaps enough to ensure that no such problems will arise.   

 Having a Hamiltonian formalism allowed us to make contact with detailed Hamiltonian results for the problem of fermions at unitarity in harmonic traps. We have seen that the conformal quantum mechanics in the holographic coordinate is hidden in the standard many-body bound state problem in terms of the effective Hamiltonian for
 the mean-square size of the trapped droplet, thus giving a direct interpretation of the holographic coordinate in the physical system of interest. On the other hand, the main computational challenge, i.e.~the determination of
 the $\Lambda_{(N,d)}$ spectrum, is simply reformulated in terms of the mass spectrum of bulk fields. In the absence
 of a fundamental definition of the bulk theory (such as the type IIB string theory in the case of $N=4$ SYM), the holographic picture does not yet help in determining the $\Lambda_{(N,d)}$ coefficients. Perhaps one could contemplate a background containing an explicit factor of the ${\bf S}^{dN-d-1}$ sphere, modded by an appropriate discrete subgroup of $O(dN-d)$, in such a way that the $\Lambda_{(N,d)}$ coefficients arise as direct Kaluza--Klein contributions to scalar field masses. 
 
 At a more heuristic level, we have considered possible patterns for mass gap generation  by looking at particular deformed metrics where only the time-time component is affected. This is enough to generate quite arbitrary potentials in the holographic coordinate, capable of discretizing the spectrum of excitations.    We have estimated the contribution of such towers of quasiparticles to the thermal entropy and concluded that they mimic extra dimensions. This is similar to the graviton gas contribution to the entropy of strongly coupled CFTs in the standard ${\rm AdS}_5 \times {\bf S}^5$ model, giving
a band with entropy $S(E)_{g} \sim (ER)^{9/10}$. We know that this changes into a black-hole dominated
entropy at  energies in excess of $N^2 /R$, where $N^2$ is the number of `gluon partons', into a law $S(E)_{bh} \sim N^2 (ER)^{3/4}$, i.e.~like a four-dimensional plasma. In the canonical ensemble, the competition between these two bands of states produces the Hawking--Page transition at 
$T_c \sim 1/R$. 

It is very interesting to ask what similar pattern would arise in the nonrelativistic models at hand in this paper. The exactly analogous set up to the relativistic CFT on a sphere is the model with an $SL(2, \mathds{R})$ harmonic trap. The full single particle spectrum is given by
\beq\label{se}
\omega_{{\vec n}, q} = \Omega \left( \sum_{i=1}^d n_i + 2q + \nu+ {d+2 \over 2}\right)
 \;.\eeq
 The associated free energy of the boson gas,
 \beq\label{trf}
 \beta F(\beta) = \sum_{({\vec n}, q)\in \mathds{Z}_+^{d+1}} \log\left(1-e^{-\beta \omega_{{\vec n}, q} }\right)\;,
 \eeq
 yields a  high-temperature scaling of the entropy
 $S(T) \sim (T/\Omega)^{d+1}$, corresponding to one extra dimension. If the analog of a Hawking--Page transition would take place, we expect $T_c \sim \Omega$ and a phase at $T>T_c$ well described by the `partonic' degrees of freedom. If the bulk field $\phi$ is dual to some composite operator of the form 
 $\sum_{a=1}^{N_f} \psi_a \psi^a$ we have $N_f$ partons. The entropy of such gas in a harmonic trap
 is 
 \beq\label{partt}
 S(T)_{\rm partons} \sim N_f (T/\Omega)^{d}\;
 \eeq
 at high temperature. This means that a Hawking--Page transition occurring at $T=T_c \sim \Omega$ would have a latent
 heat of order $T_c \Delta S \sim \Omega N_f$, which diverges in the large $N_f$ limit. Hence, such a transition would be first order and perhaps described by the nucleation of a black hole, just like 
 in the relativistic case. This  would require introducing dynamical gravity into the picture and
 completing the parameter dictionary so that the Planck length is related to the AdS curvature by the relation
 \beq\label{lastdi}
 {R^{d+1} \over G_{\rm N}} \sim \left( {R \over \ell_P}\right)^{d+1} \sim N_f
 \;.\eeq
 In this case, gravitational interaction vertices in the bulk scale naturally as inverse powers of $N_f$ since
 the effective low-energy expansion parameter  would become $G_N /R^{d+1} \sim 1/N_f$. 
 
 Finally, true deconfining transitions would occur in the models described at the end of the last section, 
 when the quasiparticles, or `molecules', dissociate into the  $N_f$ partonic `atoms'.  For example, in the model based on a sharp wall, with internal energy gap $\Delta \varepsilon \sim 1/ML_\rho^2$. We can expect the critical dissociation  temperature to be of this order, $T_{\rm dec} \sim \Delta \varepsilon$. The
 entropy of the molecular gas at this temperature is given by Eq. ({\ref{scalen}),
 \beq\label{critent}
 S_{\rm molecules} \sim VL_\rho (MT_{\rm dec})^{d+1 \over 2} \sim V/L_\rho^d \;.
 \eeq
 On the other hand, the entropy of an `atomic' gas with $N_f$ species is 
 \beq\label{pare}
 S_{\rm atoms} \sim  N_f V (MT_{\rm dec})^{d/2} \sim  N_f V/L_\rho^d\;.
 \eeq
 Again, we find a latent heat of $O(N_f)$ and we expect again a black-hole type description for this
 transition, since $N_f \propto 1/G_{\rm N}$.  It should be interesting to dwell further on these questions.

 Finally, it would be important to go beyond the field-theoretical description of the bulk system,
 and search for some embedding in string theory. In such a model, irrespectively of its actual physical applications, we
 have a well defined spectrum of excitations with bulk masses dictated by the fundamental theory (for example, in such
 a model  any  matching like  Eq. (\ref{nup}) would become a prediction rather than a fit). In this endeavor, the first
 step is to search for supersymmetric generalizations of this system, for which the supersymmetric generalization of
 conformal quantum mechanics, studied in \cite{rabino}, will certainly be crucial. See also \cite{sakaguchi}.
 
\vspace{0.2cm}
{\bf Acknowledgments}

\vspace{0.2cm}

We would like to thank David  B. Kaplan for an enlightening lecture on  delta function potentials and dimensional transmutation. 
This work was partially supported by Spanish Research Ministry, MEC and FEDER
under grant FPA2006-05485,  Spanish Consolider-Ingenio 2010 
Programme CPAN (CSD 2007-00042), Comunidad Aut\'onoma de Madrid, CAM under grant HEPHACOS P-ESP-00346, Spanish Consolider-Ingenio 2010, and  the European Union Marie Curie RTN network under
contract MRTN-CT-2004-005104. C.A.F. enjoys a FPU fellowship from MEC under grant AP2005-0134.


\end{document}